# *HighProbability* determines which alternative hypotheses are sufficiently probable

## Genomic applications include detection of differential gene expression

**David R. Bickel**

*Office of Biostatistics and Bioinformatics*
*Medical College of Georgia*
*Augusta, GA 30912-4900*
*bickel@prueba.info*
*www.davidbickel.com*

February 27, 2004

**Abstract:** Many genomic experiments, notably microarray experiments seeking to detect differential gene expression, involve calculating a large number of p-values. This leads to the *multiple testing problem*: when the number of null hypotheses is large, the probability of accepting at least one false alternative hypothesis is often much greater than the significance level of the tests, which tends to mislead investigators. Software called HighProbability provides a simple, fast, reliable solution to the multiple testing problem, with applications to many areas of bioinformatics. For example, in a microarray study, HighProbability can determine which genes are probably differentially expressed. Given a set of p-values not adjusted for multiple testing, HighProbability determines which ones are low enough to imply a high probability of the truth of their alternative hypotheses. The set of p-values may be determined by conventional hypothesis testing or by random permutations using existing R or S-PLUS software. HighProbability is freely available under license through http://www.davidbickel.com. Coded in S, HighProbability currently requires an installation of R or S-PLUS, but the algorithm is short enough for fast implementation in non-S languages as well.

## INTRODUCTION

Many problems in bioinformatics involve testing a large number of null hypotheses. In statistical terminology, each null hypothesis is true if and only if an alternative hypothesis is false; each null hypothesis is a statement that there is no effect in the population, so its corresponding alternative hypothesis states that there is an effect in the population. For example, in a two-group microarray study, the alternative hypothesis associated with a gene may state that it is differentially expressed across the two populations from which the two groups are considered random samples. Conventional hypothesis testing assigns a p-value to each test such that, for a given test, the null hypothesis is rejected (considered false) and the alternative hypothesis is accepted (considered true) only if the p-value is less than or equal to some significance level $\alpha$ such as $\alpha$=0.05. Then, if the null hypothesis is true, $\alpha$ is the probability of making a Type I error, i.e., of accepting the false alternative hypothesis. ($\alpha$ is commonly, but completely, misunderstood as the probability that the null hypothesis is true.) The multiple testing problem is that when $m$, the number of hypotheses, is large, the probability of accepting at least one false alternative hypothesis is much greater than $\alpha$, often misleading investigators. A powerful, flexible approach to the multiple testing problem is to find the value of $\alpha$ such that the probability that an accepted alternative hypothesis is true equals or exceeds some specified level $p_1$. This level depends on the goals of the investigator: Efron *et al.* (2001) used $p_1 = 90\%$ to ensure a high confidence in accepted alternative hypotheses, and Kendziorski *et al.* (2003) used $p_1 = 50\%$ to determine which alternative hypotheses are more likely than not. The approach of the former relied on nonparametric density estimation, whereas that of the latter relied on parametric modeling, but the value of $p_1$ is not tied to the approach.

It has been shown that this problem of determining which alternative hypotheses have posterior probabilities of at least $p_1$ is equivalent to the problem of deciding which alternative hypotheses are most advantageously considered true based on $(1 - p_1)^{-1} - 1$ as the ratio of the cost of considering a false alternative true to the benefit of considering a true alternative true (Bickel 2004b). (Müller *et al.* (2004) noticed an equivalent relationship between decision-theoretic optimization and using a posterior probability threshold.) Unlike other posterior probability threshold methods, the method of Bickel (2004b) does not require the estimation of any probability density. A nonparametric form of the method performs much better than a related nonparametric method based on density estimation (Bickel 2004b). Nonparametric methods have the advantage of obviating distributional assumptions and extensive model validation. Thus, the decision-theoretic method of determining which alternative hypotheses have probabilities above some threshold (Bickel 2004b) has been implemented as HighProbability to make it readily available to the scientific community.

## IMPLEMENTATION & ALGORITHM

The current implementation of HighProbability requires either of the two S engines, R or S-PLUS. Each engine has its advantages; most notably, R (R Development Core Team, 2003) is freely available and has convenient scoping rules, whereas S-PLUS (Insightful Corp.) comes with a high-quality graphical user interface and telephone technical support. Software suites implementing some of the latest statistical methods for microarray analysis are the Bioconductor.org suite for R and the S+ArrayAnalyzer suite (Insightful Corp.) for S-PLUS. However, these add-on suites are not needed for HighProbability since R and S-PLUS have many reliable built-in functions for the computation of p-values.

The main algorithm of HighProbability is short enough to be easily implemented in non-S languages as well:

**Input:**

- Required: $\mathcal{P}$, a vector of p-values, $\mathcal{P}_1, \mathcal{P}_2, \ldots, \mathcal{P}_m$ (with one p-value corresponding to each of $m$ hypothesis tests);
- Optional: $p_1$, the lowest probability that an alternative hypothesis can have for it to be considered true, conditional on the p-value (this is the minimum acceptable probability of differential expression for a gene to be considered differentially expressed, and is 50% by default);
- Optional: $\pi_1$, the marginal probability that an alternative hypothesis is true. (This can be conservatively set to 0, as per Benjamini and Hochberg (1995), but the methods of Storey (2003) can yield less conservative estimates of $\pi_1 = 1 - \pi_0$, where $\pi_0$ is the marginal probability that a null hypothesis is true. By default, a recursive call to the algorithm with $\pi_1 = 0$ and $p_1 = 50\%$ is used to estimate $\pi_1$. This could be repeated until convergence, but a single iteration is sufficient for most purposes.)

**Output:**

- **I**, a vector of indicator (boolean) values, $I_1, I_2, \ldots, I_m$, corresponding to the p-values, where TRUE indicates that the corresponding alternative hypothesis has a probability of at least $p_1$, and FALSE indicates that it does not. That is, the $j$th null hypothesis is rejected and the $j$th alternative hypothesis is accepted if $I_j = \text{TRUE}$; otherwise, the $j$th null hypothesis is not rejected.

**Step 1.** Set the levels of significance, denoted by the vector $\alpha$, equal to the vector of p-values, i.e., $\forall_{j \in \{1,2,\ldots,m\}} \ \alpha_j \leftarrow \mathcal{P}_j$. Each level of significance is a p-value threshold and a test-wise Type I error rate.

**Step 2.** For $j = 1, 2, \ldots, m$, at the $j$th level of significance, compute, $R_j$, the number of rejections of the null hypothesis. This is the number of p-values less than or equal to the $j$th significance level: $R_j \leftarrow \#_{k \in \{1,2,\ldots,m\}} \ \mathcal{P}_k \le \alpha_j$.

**Step 3.** For $j = 1, 2, \ldots, m$, estimate the decisive false discovery rate (dFDR) at the $j$th level of significance using the standard estimator:
$$\hat{\Delta}_j \leftarrow \begin{array}{ll} \frac{(1-\pi_1)\,\alpha_j}{R_j/m}, & R_j > 0 \\ 0, & R_j = 0 \end{array}.$$

**Step 4.** For $j = 1, 2, \ldots, m$, compute the relative net desirability or relative gain with $\alpha_j$ as the significance level (Bickel, 2003): $g_j \leftarrow \left(1 - \frac{\hat{\Delta}_j}{1-p_1}\right) R_j$.

**Step 5.** Compute $\alpha_{\text{optimal}}$, the significance level at which the relative gain is maximized: $j(\text{optimal}) \leftarrow \text{argmax}_{j \in \{1,2,\ldots,m\}} \ g_j$; $\alpha_{\text{optimal}} \leftarrow \alpha_{j(\text{optimal})}$. (Notice that $g_{j(\text{optimal})} = \max_{j \in \{1,2,\ldots,m\}} g_j$.)

**Step 6.** For $j = 1, 2, \ldots, m$, consider the $j$th alternative hypothesis sufficiently probable if and only if it is accepted at significance level $\alpha_{\text{optimal}}$ and $g_{j(\text{optimal})} > 0$:
$$I_j \leftarrow \begin{array}{ll} \text{TRUE}, & \mathcal{P}_j \le \alpha_{\text{optimal}} \text{ and } g_{j(\text{optimal})} > 0 \\ \text{FALSE}, & \mathcal{P}_j > \alpha_{\text{optimal}} \text{ or } g_{j(\text{optimal})} \le 0 \end{array}.$$

For readers more familiar with S than with mathematical notation, the algorithm may be better understood by an examination of the HighProbability function `alternative.probable`, which prints on less than two pages. The source code also specifies the details of handling errors, ties, and special situations.

## DEMONSTRATION

The main function of HighProbability is `alternative.probable`, illustrated here. The function `alternative.beneficial` does the same thing, except that in replacing the $p_1$ argument with a cost-to-benefit ratio argument, it implements Bickel (2003). Although HighProbability applies to general multiple testing problems, the special terminology of differential expression will make an important application to microarrays clear. Consider 10 simulated microarrays, each with expression levels of 1000 genes of interest. The logarithms of the expression levels (LELs) were independently and randomly generated from the normal distribution with zero mean and unit variance, except for those of the first 100 genes of the five microarrays constituting the first group, which were generated from the normal distribution with a mean of two and a variance of one. Thus, only the first 100 genes are differentially expressed between that group and the remaining five microarrays, which constitute the second group. As if that were not known, a p-value was computed for each of the 1000 genes using the two-sided, equal-variance t-test. (Various t-tests, Wilcoxon rank-sum tests, and nonparametric permutation tests are easily performed using the `t.test`, `wilcox.test`, and `sample` functions of S, respectively. The `wilcox.test` function facilitated nonparametrically testing more biologically relevant null hypotheses (Bickel 2004a).) These 1000 p-values were stored in the vector `p.values`, the actual argument of `alternative.probable`:

```
> ap.50 <- alternative.probable(p.values)
Using marginal.probability estimate of  0.088
```

Here, $p_1 = 50\,\%$, as no value was specified for the formal argument `min.probability`. Likewise, since the argument `marginal.probability`, corresponding to $\pi_1$, was not be specified, $\pi_1$ was estimated by initially using the most conservative value, `marginal.probability = 0`. As desired, the estimate, 8.8%, is close to but no greater than the actual value, 10%, which would be unknown. The values of the indicator vector **I** are stored in `ap.50`. HighProbability discovered 60 of the 100 differentially expressed genes and falsely discovered only 28 of the 900 equivalently expressed genes:

```
> sum(ap.50[1:100])
[1] 60
> sum(ap.50[101:1000])
[1] 28
```

To reduce the number of false discoveries, but also the number of true discoveries, the investigator can use $p_1 = 90\%$:

```
> ap.90 <- alternative.probable(p.values, min.probability = 0.90)
Using marginal.probability estimate of  0.088
> sum(ap.90[1:100])
[1] 13
> sum(ap.90[101:1000])
[1] 2
```

In this case, the results are not affected by using 0% rather than 8.8% as the value of $\pi_1$:

```
> ap.conservative.90 <- alternative.probable(p.values, min.probability = 0.90,
marginal.probability = 0)
> sum(ap.conservative.90[1:100])
[1] 13
> sum(ap.conservative.90[101:1000])
[1] 2
```

As time and computation power allow, investigators may perform exploratory analysis with HighProbability by varying $p_1$ from 50% to 95% in increments of 5%. Finer increments can be used to estimate the probability of differential expression for each gene (Bickel 2004b). When set to TRUE, the additional argument `plot.relative.gain` produces a plot of $g_j$ versus $\alpha_j$:

```
> ap.50 <- alternative.probable(p.values, plot.relative.gain = TRUE)
```

$\alpha_{\text{optimal}}$ is the value of the horizontal axis at which $g_j$ reaches its maximum.